\documentclass[12pt,tightenlines,eqsecnum,floats,aps,amsmath,amssymb,nofootinbib,superscriptaddress,showpacs]{revtex4}
\usepackage{dcolumn}
\usepackage{bm}
\usepackage[latin1]{inputenc}
\usepackage[spanish,english]{babel}
\usepackage{amsfonts}
\usepackage{amssymb}
\usepackage{graphicx}

\newcommand{\be}{\begin{equation}}
\newcommand{\ee}{\end{equation}}
\newcommand{\bea}{\begin{eqnarray}}
\newcommand{\eea}{\end{eqnarray}}

\begin{document}

\title{{\bf Acceleration radiation, transition probabilities, and trans-Planckian physics }}

\author{Ivan Agullo}\email{ivan.agullo@uv.es}
 \affiliation{ {\footnotesize Physics Department, University of
Wisconsin-Milwaukee, P.O.Box 413, Milwaukee, WI 53201 USA}}
\author{José Navarro-Salas}\email{jnavarro@ific.uv.es}
\affiliation{ {\footnotesize Departamento de Física Teórica and
IFIC, Centro Mixto Universidad de Valencia-CSIC.
    Facultad de Física, Universidad de Valencia,
        Burjassot-46100, Valencia, Spain. }}

\author{Gonzalo J. Olmo}\email{olmo@iem.cfmac.csic.es }
\affiliation{{\footnotesize Instituto de Estructura de la Materia,
CSIC, Serrano 121, 28006 Madrid, Spain}}\affiliation{ {\footnotesize Physics Department, University of
Wisconsin-Milwaukee, P.O.Box 413, Milwaukee, WI 53201 USA}}

\author{Leonard Parker}\email{leonard@uwm.edu}
\affiliation{ {\footnotesize Physics Department, University of
Wisconsin-Milwaukee, P.O.Box 413, Milwaukee, WI 53201 USA}}

\begin{abstract}

An important question in the derivation of the acceleration radiation, which also arises in Hawking's derivation of black hole radiance, is the need to invoke trans-Planckian physics for the quantum field
that originates the created quanta. We point out that this issue can be  further clarified by reconsidering the analysis in terms  of  particle detectors, transition probabilities, and local two-point functions. By writing down separate expressions for the spontaneous- and induced-transition probabilities of a uniformly accelerated detector, we show that the bulk of the effect comes from the natural (non trans-Planckian) scale of the problem, which largely diminishes the importance of the trans-Planckian sector. This is so, at least, when trans-Planckian physics is defined in a  Lorentz invariant way. This analysis also suggests how to define and estimate the role of trans-Planckian physics in the Hawking effect itself.\\

\end{abstract}


\maketitle

\section{Introduction}
After formulating general relativity Einstein returned to the microscopic world. He introduced the concept of transition probabilities between stationary states in the context of the interaction of atoms with radiation. He established a link between black-body radiation and the theory of atomic spectra. In short, Einstein considered transitions between two states, an upper excited state $2$ and a lower state $1$, with energies $E_2>E_1$. The probability per atom and per unit time of a jump from state $1$ to state $2$ induced by the environment radiation is
\be \label{Eif}\dot P_{1\to 2}= Bu_{w} \ , \ee where $u_{w}$ is the energy density of radiation at the frequency $w= (E_2-E_1)/\hbar$ and $B$ is one of the so called Einstein coefficients.
In addition, the probability per atom and per unit time for the decay of the state $2$ to state $1$ is assumed to be
\be \label{Efi}\dot P_{2\to 1}= Bu_{w} + A\ , \ee
where $A$ represents the probability of spontaneous emission and $Bu_{w}$ gives  also the probability of induced emission.  Thermal equilibrium is achieved if
\be N_1\dot P_{1\to 2}=N_2\dot P_{2\to 1} \ , \ee
when the  state population quotient $N_2/N_1$ obeys the Boltzmann distribution for probabilities  $N_2/N_1= e^{-\Delta E/k_BT}$ at the equilibrium temperature $T$ ($k_B$ is the Boltzmann constant and $\Delta E\equiv E_2-E_1$). Einstein realized that thermal equilibrium implies that $u_w$  turns out to be the Planck law for the energy  density, provided that the quotient $A/B$ is just $A/B=2 \hbar w^3/\pi c^3$. This analysis can be used to infer the thermal character of the environment radiation by analyzing only the transition probabilities of the atomic system; the environment radiation is thermal provided that the transition probabilities between the energy levels of the atomic system satisfy, at equilibrium, the so called detailed balance relation $\dot P_{1\to 2}/\dot P_{2\to 1}= e^{-\Delta E/k_BT}$.

Many years later, physicists working in the theory of quantum fields in curved space  realized that an atomic system following a uniformly accelerated worldline in Minkowski spacetime, with acceleration $a$, feels itself immersed in a thermal bath at the temperature $T=\hbar a/2\pi c k_B$, when the quantum state of the field is the ordinary Minkowski vacuum.
The acceleration radiation effect can be analyzed from two different points of view. It can be derived by computing the expectation value of the number operator in the Minkowski vacuum state by using the formalism of Bogolubov transformations \cite{parker} in Rindler space \cite{fulling, davies}. The Bogolubov coefficient approach is also the basis of  Hawking's original derivation of black hole radiance \cite{hawking} (see also \cite{parker-toms}). On the other hand, the acceleration radiation effect can also be derived by studying the transition rate probabilities of  uniformly accelerated particle detectors in Minkowski spacetime \cite{unruh} (see also the review \cite{crispino}). In this approach, the transition probabilities are often written in terms of the two-point function of the Minkowski vacuum state. In this form, the derivation is somewhat closer to the derivation of black hole radiance carried out by Fredenhagen and Haag \cite{fredenhagen-haag}.

When dealing with the acceleration radiation or the Hawking effect, an important question arises. To what extent are these thermal effects sensitive to trans-Planckian physics? In
Hawking's original derivation this issue emerges naturally because emitted quanta reaching future null infinity at sufficiently late times
suffer an arbitrarily large blueshift when propagated backwards
in time to past null infinity. In fact, the precursors of the Hawking quanta can have trans-Planckian frequencies in the vicinity of the horizon (see for instance \cite{Parker77, Wald84, jacobson9193}).\footnote{This issue has been traditionally addressed by explicit modification of the standard relativistic dispersion relations \cite{dispersionrelations}.  Here, we follow a different approach that preserves the relativistic invariance.}
The same question arises in the derivation of the acceleration radiation.  This is because, in any given inertial frame, the uniformly accelerated detector acquires an arbitrarily large velocity after sufficient proper time $\tau$ and, correspondingly, the thermal quanta it is observing at such times correspond to modes with arbitrarily large frequencies $w' \sim w e^{a\tau}$ relative  to the given inertial frame \cite{wald01}.
This fact is manifest in the derivation in terms of Bogolubov coefficients, which requires an unbounded integral in frequencies in the intermediate steps of the derivation.  However, these modes are not detected by an inertial observer and their physical relevance is not clear. On the other hand, in the derivation of the acceleration radiation in terms of two-point functions, trans-Planckian physics seems to appear because ultra-short lapses of proper time are apparently important in obtaining the final result. However, this inference depends on the distributional character of the two-point function.

In this paper we reanalyze this problem by studying  the transition probabilities of uniformly accelerated particle detectors. We parallel Einstein's analysis by computing separately the induced and spontaneous transition probabilities of the detector and we obtain the thermal character of the radiation by means of the detailed balance relation. The splitting of the different contributions has the advantage of providing suitable mathematical expressions that allow us to define and evaluate the contribution of trans-Planckian physics in a Lorentz invariant way.
We find that the thermal outcome arises from scales of the same order as the acceleration $a$ itself, which strongly suggests that the  effect is indeed a low-energy phenomenon.

In section II, we review the standard analysis of the acceleration radiation in Rindler spacetime in terms of Bogolubov coefficients. In section III, we compute the spontaneous and induced transition probability rates of a uniformly accelerated particle detector in Minkowski spacetime, and we use the results to show that the radiation felt by the detector is thermal. In section IV, we repeat this analysis using the two-point function of the quantum field. Finally, in section V, we use the results presented in sections III and IV to analyze the role of trans-Planckian physics in the computation of the acceleration radiation. There, we also make some comments regarding the same topic in Hawking radiation.  In section VI, we give our conclusions.\footnote{In the rest of the paper, we use units such that $\hbar=c=1$ and $k_B=1$.}

\section{Acceleration radiation and Bogolubov transformations \label{sec:Rindler}}

The acceleration radiation was first derived in the context of the formalism of Bogolubov transformations relating inertial and accelerated modes. In this section we will quickly review this derivation. \\

A uniformly accelerated (Rindler) observer has a natural coordinate system $(\tau, \xi, y, z)$ related with the inertial coordinates $(t,x,y,z)$  by
\begin{equation} \label{accelerated2}
t =\frac{e^{a\xi}}{a}
\sinh{a \tau} \ , \
x=\frac{e^{a\xi}}{a} \cosh{a \tau} \ , \ y=y \ , \ z=z
\ . \end{equation}
The curve $\xi=0$ represents a  uniformly accelerated trajectory with proper acceleration $a$.
 The wave equation for a massless scalar field $\Box \phi(x)=0$ in the
coordinates of the accelerated observer becomes \be
(e^{-2a\xi}(-\partial^2_{\tau}+\partial^2_\xi)+\partial^2_y+\partial^2_z
)\phi({\tau},\xi,y,z)=0\ee The $y,z$ dependence can be trivially
integrated using  plane waves $\phi(t,\xi,y,z)=\phi(t,\xi) e^{i k_y
y} e^{i k_z z}$. Introducing this ansatz in the equation, we find
\be \label{eq:waveq}
[(-\partial^2_{\tau}+\partial^2_{\xi})-e^{2a\xi}(k_y^2+k_z^2)]\phi({\tau},\xi)=0
\ . \ee This equation indicates that the free scalar field observed by the
Minkowski observer appears to the uniformly accelerated
observer like a scalar field in a repulsive
potential $V(\xi) \propto e^{2a\xi}\vec{k}_{\bot }^2$, where
$\vec{k}_{\bot}^2 = k_y^2+k_z^2$. The exact form of the normalized modes, with natural
support on the accessible region for the accelerated observer
(right-hand Rindler wedge), can be expressed as
\begin{equation} \label{modesR}
u^{R}_{w,\vec{k}_{\bot}}=\frac{e^{-iw{\tau}}}{2\pi^2\sqrt{a}}\sinh^{\frac{1}{2}}\left(\frac{\pi
w}{a}\right)K_{iw/a}\left(\frac{|\vec{k}_{\bot
}|}{a}e^{a\xi}\right)e^{i\vec{k}_{\bot}\cdot\vec{x}_{\bot}} \ ,
\end{equation}  where $\vec{k}_{\bot}\cdot\vec{x}_{\bot} =k_y y+k_z z$. The important point is that the above positive frequency modes
cannot be expanded in terms of the standard purely positive
frequency modes of the inertial observer \be
u^{M}_{k_x,\vec{k}_{\bot}}
=\frac{1}{\sqrt{2(2\pi)^3k_0}}e^{-ik_0 t+i(k_x x +
\vec{k}_{\bot}\cdot\vec{x}_{\bot})} \ , \ee where $k_0= \sqrt{k_x^2
+ \vec{k}_{\bot}^2}.$ The detailed analysis requires one to compute
the corresponding Bogolubov coefficients. They are found to be
\cite{fulling, crispino} \be \beta_{w \vec{k}_{\bot}, k'_x
\vec{k}'_{\bot}}=-\left [2\pi ak'_0(e^{2\pi w/a}-1)\right ]^{-1/2}
\left (\frac{k'_0 + k'_x}{k'_0 - k'_x} \right )^{-iw/2a}\delta
(\vec{k}_{\bot}-\vec{k}'_{\bot}) \ . \ee With this result one can compute the important physical result that follows. The mean number $n_w$ of Rindler
particles in the Minkowski vacuum is directly tied to the integral \be
n_w=\int_{-\infty}^{+\infty}d\vec{k}' \beta_{w_1 \vec{k}_{\bot},
\vec{k}'} \beta_{w_2 \vec{k}_{\bot}, \vec{k}'}^* \ . \ee The
integration in $\vec{k}_{\bot}$ is trivial and the integration in $k'_x$ reduces to \be
\label{integralkX}\int_{-\infty}^{+\infty}dk'_x (2\pi ak'_0)^{-1}
\left (\frac{k'_0 + k'_x}{k'_0 - k'_x} \right
)^{-i(w_1-w_2)/2a} \ , \ee
which, changing the integration variable to the rapidity $y=\tanh^{-1}(k_x'/k'_0)$, leads to
\begin{equation}
\label{integralkY}\int_{-\infty}^{+\infty}\frac{dy}{2\pi a}e^{-i(w_1-w_2)y/a}
= \delta(w_1-w_2) \ .
\end{equation}
Taking into
account the remaining terms, one easily gets  \be
\int_{-\infty}^{+\infty}d\vec{k}' \beta_{w_1 \vec{k}_{\bot 1},
\vec{k}'} \beta_{w_2 \vec{k}_{\bot 2}, \vec{k}'}^* =
\frac{1}{e^{2\pi w_1/a}-1} \delta (w_1 - w_2) \delta (\vec{k}_{\bot
1}-\vec{k}_{\bot 2})\ . \ee The final outcome becomes then extremely
simple and important, and parallels the Hawking effect on black hole radiance. The Minkowski vacuum can be described, in the spacetime region accessible for a uniformly accelerated observer (the Rindler wedge), as a thermal bath of Rindler quanta at temperature $ T = a/2\pi$. This result \cite{fulling, davies}  was strongly reinforced by Unruh's  interpretation in terms of particle detectors \cite{unruh}. A uniformly accelerated particle detector is excited by the absorption of a Rindler quantum from the thermal bath. An inertial observer describes this process in a different way, as the emission of a Minkowski particle as the result of the interaction of the detector with the quantum field \cite{unruh-wald}, as explicitly worked out in next section.

\section{Transition probabilities of an accelerated particle detector \label{sec:III}}

In this section we review the particle detector approach to the acceleration radiation effect. We compute separately the spontaneous and induced emission and absorption processes. The thermal character of the Minkowskian vacuum with respect to an accelerated observer is derived via the detailed balance relation.

 \subsection{Spontaneous emission of a uniformly accelerated detector}

 Let us consider a quantum mechanical system coupled to a free massless scalar quantum field $\Phi(x)$ in Minkowski spacetime. For simplicity the field is assumed massless. The quantum mechanical system modeling our particle detector \cite{unruh, dewitt} will have some internal energy states $|E\rangle$, which are eigenstates of the corresponding free Hamiltonian $H_q$. We will consider here two of those sates,  $|E_2\rangle$ and $|E_1\rangle$, with energies $E_2>E_1$.  The detector can interact with the quantum field by absorbing or emitting  quanta.
The interaction can be modeled in a simple way by coupling the field $\Phi(x)$ along the detector's trajectory $x=x(\tau)$ ($\tau$ is the detector's proper time) to a monopole moment operator $m(\tau)$ acting on the internal detector eigenstates through the Lagrangian
\be \label{detectorPhi} {L}_I=  g \ m(\tau)\Phi(x(\tau)) \ , \ee
where $g$ is the strength of the coupling. In the interaction picture the detector's operator $m(\tau)$ has the standard unitary time evolution $m(\tau)= e^{iH_q \tau}m(0)e^{-iH_q \tau}$.

Before analyzing  the accelerated detector, it is useful to consider the simple example of an inertial detector. The spontaneous emission of an inertial detector can be studied by considering the transition amplitude for the process $|E_2\rangle |0_M\rangle \  \to  |E_1\rangle |\psi\rangle $, where  $|0_M\rangle$ is the usual Minkowski vacuum state and  $|\psi\rangle$ is the final state of the field. The field $\Phi(x)$ can be  quantized by expanding it in standard plane-wave modes
\be \Phi(x)= \int d^3k \left(u^M_{\vec{k}} a_{\vec{k}} + u^{M*}_{\vec{k}} a^{\dagger}_{\vec{k}}\right) \ , \ee
with
\be u^M_{\vec{k}}= \frac{1}{\sqrt{2(2\pi)^3w}} e^{-i(wt-\vec{k}\vec{x})} \ , \ee
where $t$ and $x$ are inertial coordinates and $w=|\vec{k}|$. The amplitude for the process is given, to first order in time-dependent perturbation theory, by
\be
 ig \langle E_1|m(0)| E_2\rangle \int d\tau e^{i(E_1-E_2) \tau} \langle \psi|\Phi(x(\tau))|0_M\rangle \ . \ee
Because of the monopolar interaction, this transition can only take place to one-particle (Minkowski) states. Taking $|\psi \rangle = |\vec{k}\rangle$, the corresponding amplitude is then\footnote{Note that one could choose instead of $|\psi \rangle = |\vec{k}\rangle$ a superposition of one-particle states. However, since at the end we sum over all possible final states, the outcome will be independent of the particular basis chosen. Our choice is thus made on grounds of technical and notational simplicity.}
\be  ig \langle E_1|m(0)| E_2\rangle \int d\tau e^{i(E_1-E_2) \tau} \frac{1}{\sqrt{2(2\pi)^3w}} e^{i(wt(\tau)-\vec{k}\vec{x}(\tau))} \ , \ee
where $(t(\tau),\vec{x}(\tau))$ is the trajectory of the detector. For the inertial detector we have $t=\tau, \vec{x}=0$. The transition probability  to the final state $|E_1\rangle  |\vec{k}\rangle$ is then given by squaring the above expression

\be P_{2\to 1, \vec{k}}= g^2 |\langle E_1|m(0)| E_2\rangle|^2 \frac{1}{2(2\pi)^3w}\int d\tau_1d\tau_2 e^{i(E_1-E_2+ w) (\tau_1 - \tau_2)} \ , \ee
where $w=|\vec{k}|$.
Therefore, the corresponding transition probability per unit time is then given by ($\Delta \tau\equiv \tau_1-\tau_2$)
\bea {\dot P}_{2\to 1, \vec{k}}&=& g^2 |\langle E_1|m(0)| E_2\rangle|^2 \frac{1}{2(2\pi)^3w}\int d\Delta \tau e^{-i(\Delta E- w) \Delta \tau}\nonumber \\ &=& g^2 |\langle E_1|m(0)| E_2\rangle|^2 \frac{1}{2(2\pi)^2w}\delta (\Delta E - w)\ , \eea
where the delta function reflects the energy conservation of the process, with $\Delta E \equiv E_2 -E_1>0$. The transition $E_2 \to E_1$  is accompanied by the emission of a quantum of the field with energy $w=\Delta E$.
Finally, the total transition probability rate for the detector  is obtained by summing over all possible one-particle final states
\footnote{Had we chosen a non-static inertial observer with $\vec{x}=\vec{v}t$, the delta function would take the form $\delta(\Delta E-\gamma (w-\vec{k} \cdot\vec{v}))$, but the final result  is the same as for $\vec{v}=0$.}

\bea \label{A21f} \dot P_{2\to 1}({\rm spontaneous})&=& g^2 |\langle E_1|m(0)| E_2\rangle|^2 \int d\Omega_{\vec{k}}w^2 dw \frac{1}{2(2\pi)^2w}\delta (\Delta E- w)\nonumber \\ &=& g^2 |\langle E_1|m(0)| E_2\rangle|^2  \frac{\Delta E}{2\pi}\ . \eea

The spontaneous emission rate  is an intrinsic property of the detector and its interaction with the quantum field. Therefore, it is {\it insensitive} to the trajectory of the detector and is given by the previous expression. We can check this  by  computing explicitly the spontaneous emission rate of a detector following a uniformly accelerated trajectory (see appendix A for explicit derivation of the same result for a freely falling detector in de Sitter spacetime).

Let us then consider that the detector follows a uniformly accelerated trajectory with proper acceleration $a$
\begin{equation} \label{accelerated}
t =\frac{1}{a} \sinh{a\tau} \ , \ x= \frac{1}{a} \cosh{a\tau} \ , \ y=0 \ , \ z=0 \ \ .
\end{equation}
One can easily repeat the above calculation for the process $|E_2\rangle |0_R\rangle \  \to  |E_1\rangle |\psi\rangle $, where now the initial state of the quantum field, $|0_R\rangle$, is taken as vacuum associated to the uniformly accelerated observer (usually called the Rindler vacuum) and  $|\psi\rangle $ stands for the associated one-particle (Rindler) state. Using the coordinates $(\tau, \xi, y, z)$ associated to the accelerated, the modes defining the quantization  are those given in (\ref{modesR}).
On the accelerated trajectory we have $\xi=0$ and, for simplicity, we take ${\vec{x}}_{\bot}=(0,0)$. Then
\be u^R_{w, {\vec{k}}_{\bot}}(\tau) = \frac{e^{-iw\tau }}{2\pi^2\sqrt{a}}\sinh^{1/2}(\frac{\pi w}{a})K_{\frac{iw}{a}}(\frac{|{\vec{k}}_{\bot}|}{a}) \ . \ee
Using the same arguments as for the inertial detector, we can express the transition probability rate for all possible one-particle (Rindler) final states as

\bea \label{A21integral}  \dot P_{2\to 1}({\rm spontaneous})&=& g^2 |\langle E_1|m(0)| E_2\rangle|^2 \times \\ \nonumber & \times &  \int_0^{+\infty} |{\vec{k}}_{\bot}|\ d|{\vec{k}}_{\bot}| \ dw \ |K_{\frac{iw}{a}}(\frac{|{\vec{k}}_{\bot}|}{a})|^2 \frac{(2\pi)^2}{(2\pi^2\sqrt{a})^2}\sinh(\frac{\pi w}{a})\delta (E_1 -E_2 +w)\ , \eea
where a factor $2\pi$ comes from the one-dimensional angular integration of the transverse momentum. Performing the integral in $|{\vec{k}}_{\bot}|$ we have

\be \label{spontaneousrate} \dot P_{2\to 1}({\rm spontaneous})= g^2 |\langle E_1|m(0)| E_2\rangle|^2  \frac{\Delta E}{2\pi}\ , \ee that, as expected, exactly coincides with the result (\ref{A21f}).

\subsection{Induced  emission of a uniformly accelerated detector}

We study now the process of induced emission of a uniformly accelerated detector in Minkowski spacetime when the quantum field is in the usual Minkowski vacuum state. Let us consider then the process $|E_2\rangle |0_M\rangle \to |E_1\rangle |\psi\rangle$ for a uniformly accelerated trajectory. Since the initial state $|0_M\rangle$ is not the vacuum state for an accelerated observer, one would expect the transition  rate for this  process to be enhanced by induced emission. We will obtain the probability rate for the process of induced emission by computing the total emission probability rate and subtracting from it the spontaneous emission rate. As before, the only non-vanishing contribution will be for one particle Minkowski states, so we consider $|\psi \rangle = |\vec{k}\rangle$. The corresponding amplitude is then
\be ig \langle E_1|m(0)| E_2\rangle \int d\tau e^{i(E_1-E_2) \tau} \frac{1}{\sqrt{2(2\pi)^3w}} e^{iw(t(\tau)-\cos{\theta} x(\tau))} \ , \ee
where $t=t(\tau), x= x(\tau)$ is the accelerated trajectory (\ref{accelerated}) and $\theta$ is the angle between $\vec{k}$ and the $x$-axis. Taking into account the form of the trajectory (\ref{accelerated}), this amplitude can be rewritten as
\be  \frac{ig \langle E_1|m(0)| E_2\rangle}{\sqrt{2(2\pi)^3w}}  \int d\tau e^{i(E_1-E_2) \tau} e^{iw/2a (e^{a \tau}-e^{-a \tau}-\cos{\theta} (e^{a \tau}+e^{-a \tau}))} \ . \ee
Squaring the modulus of the above amplitude we get the transition probability
\bea P_{2\to 1,\vec{k}}&=& \frac{g^2 |\langle E_1|m(0)| E_2\rangle|^2}{2(2\pi)^3w}  \int d\tau_1 d\tau_2 e^{i(E_1-E_2) (\tau_1-\tau_2)} \nonumber \\ &\times& e^{iw/2a (e^{a\tau_1}-e^{-a \tau_1}-e^{a\tau_2}+e^{-a \tau_2}-\cos{\theta} (e^{a \tau_1}+e^{-a \tau_1}-e^{a \tau_2}-e^{-a \tau_2}))}\ . \eea
Defining $\Delta \tau = \tau_1 -\tau_2$ and $\Delta\tau^+ = (\tau_2 +\tau_1)/2$, we can rewrite $P_{2\to 1,\vec{k}}$ as
\bea P_{2\to 1,\vec{k}}&=& \frac{g^2 |\langle E_1|m(0)| E_2\rangle|^2}{2(2\pi)^3w}  \int d\Delta\tau^+ d\Delta\tau e^{i(E_1-E_2) \Delta\tau} \nonumber \\ &\times& e^{i2wa^{-1} \sinh{(a \Delta \tau/2)}[\cosh{(a \Delta \tau^+)}-\cos{\theta}\sinh{(a \Delta \tau^+)}]}\ . \label{eq:time_int}\eea
To work out the integral in $\Delta \tau$ it is convenient to
perform the change of variable $U\equiv a^{-1} e^{-a \Delta\tau/2}$ and capture the dependence on $\Delta\tau^+$ in the positive definite variable $\alpha=\cosh{(a \Delta \tau^+)}-\cos{\theta}\sinh{(a \Delta \tau^+)}$. We then get
\be P_{2\to 1,\vec{k}}=\frac{g^2 |\langle E_1|m(0)| E_2\rangle|^2}{2(2\pi)^3w}  \int d\Delta\tau^+   2 a^{i 2 \Delta E/a -1} \int_0^{\infty} dU U^{i 2 \Delta E/a -1} e^{-i \alpha w (U -(a^2U)^{-1})}\ . \label{eq:intU}\ee
The integral in $U$ does not converge absolutely. This is because we are working with plane-waves, instead of wave-packets, to represent the states $|\psi\rangle$. An integration over frequencies using wave-packets makes the integral convergent. Nonetheless, one can still work with plane-waves by inserting an infinitesimal negative real part to make the integral convergent. Therefore, one must add the appropriate $\pm i\epsilon$ terms to $w$ in the exponent.
Using now the identity
\be \label{bessel}\int_0^{\infty} dx x^c e^{a/x+b x} = 2 (-a)^{(1+c)/2} (-b)^{-(1+c)/2} K_{-1-c}(2 \sqrt{a b}) \ , \ee
for $Re[a]<0,Re[b]<0$, where $K$ is a modified Bessel function, we obtain
\be P_{2\to 1,\vec{k}}=\frac{g^2 |\langle E_1|m(0)| E_2\rangle|^2}{2(2\pi)^3w}  \int d\Delta\tau^+ \frac{4}{a} (iw-\epsilon)^{i \Delta E/a}(-iw-\epsilon)^{-i \Delta E/a} K_{-i 2 \Delta E/a}(2 w\alpha/a) \ . \ee
Taking into account that, in the limit $\epsilon\to 0^+$, $\ln(-w+i\epsilon)= i\pi + \ln w $, we get
\be P_{2\to 1,\vec{k}}=\frac{g^2 |\langle E_1|m(0)| E_2\rangle|^24e^{\pi \Delta E/a}}{2(2\pi)^3wa}  \int d\Delta\tau^+   K_{-i 2 \Delta E/a}(2 w\alpha/a) \ . \ee
The transition probability rate for this process is then given by
 \be \label{rate21k}{\dot P}_{2\to 1,\vec{k}}=\frac{g^2 |\langle E_1|m(0)| E_2\rangle|^24e^{\pi \Delta E/a}}{2(2\pi)^3wa}    K_{-i 2 \Delta E/a}(2 w\alpha/a) \ . \ee
 Summing now over all possible energies for the one-particle final states we get
\bea \label{crucialintegral}\int_0^{\infty} dw w^2{\dot P}_{2\to 1,\vec{k}}&=&\frac{g^2 |\langle E_1|m(0)| E_2\rangle|^24e^{\pi \Delta E/a}}{2(2\pi)^3a}\int_0^{\infty}dw w     K_{-i 2 \Delta E/a}(2 w\alpha/a)\nonumber \\
&=&  \frac{g^2 |\langle E_1|m(0)| E_2\rangle|^2}{4 \pi} \frac{\Delta E}{ 2 \pi}\frac{e^{2\pi \Delta E/a}}{e^{2\pi \Delta E/a}-1}\frac{1}{\alpha^2}\ , \eea
where we have made use of the following identity
\be \int_0^{\infty} dx x K_{-i a}(b x)= \frac{a \pi}{2 b^2 \sinh{(a\pi/2)}} , \ee where  $a$ and $b>0$ are real numbers. We still have to perform the angular integration. Using that
\be \int d\Omega_{\vec{k}} \alpha^{-2} = 2\pi \int_{-1}^{1} d(\cos{\theta})\frac{1}{(\cosh{(a \Delta \tau^+)}-\cos{\theta}\sinh{(a \Delta \tau^+)})^2} =4\pi \ , \ee we finally get
\be {\dot P}_{2\to 1}\equiv \int_0^{\infty}d\Omega_{\vec{k}} dw w^2{\dot P}_{2\to 1,\vec{k}}= g^2 |\langle E_1|m(0)| E_2\rangle|^2\frac{\Delta E}{2\pi}\frac{e^{2\pi \Delta E/a}}{e^{2\pi \Delta E/a}-1} \ . \label{eq:P21total} \ee
Note that, in the limit $a\to 0$ we recover the expression for the spontaneous emission, which indicates that that contribution is already contained in (\ref{eq:P21total}). Since the probability rate ${\dot P}_{2\to 1}$ is the sum of the probability rate for the spontaneous process ${\dot P}_{2\to 1}({\rm spontaneous})$ plus that of the stimulated process ${\dot P}_{2\to 1}({\rm induced})$, by subtracting ${\dot P}_{2\to 1}({\rm spontaneous})$  from (\ref{eq:P21total}) we obtain
\be \label{inducedrate}{\dot P}_{2\to 1}({\rm induced})= g^2 |\langle E_1|m(0)| E_2\rangle|^2\frac{\Delta E}{2\pi}\frac{1}{e^{2\pi \Delta E/a}-1} \ . \ee

\subsection{Absorption of a uniformly accelerated detector}

We can also consider the probability rate for the excitation of the detector $ |E_1\rangle |0_M\rangle \to |E_2\rangle |\vec{k}\rangle $, accompanied by the emission of a Minkowski particle. This can be easily extracted from (\ref{rate21k}), and  one gets
\be \label{rate12k}{\dot P}_{1\to 2,\vec{k}}=\frac{g^2 |\langle E_1|m(0)| E_2\rangle|^24e^{-\pi \Delta E/a}}{2(2\pi)^3wa}    K_{i 2 \Delta E/a}(2 w\alpha/a) \ . \ee
Summing on all possible final states one obtains the excitation probability rate ${\dot P}_{1\to 2}$
\be {\dot P}_{1\to 2}= g^2 |\langle E_1|m(0)| E_2\rangle|^2\frac{\Delta E}{2\pi}\frac{1}{e^{2\pi \Delta E/a}-1} \ , \ee
which, as expected (see (\ref{Eif}) and (\ref{Efi})), coincides with the above induced emission rate ${\dot P}_{2\to 1}({\rm induced})$.

\subsection{Thermality}

From the previous calculations we find
\be \label{eq:thermal} \frac{\dot P_{1\to 2}}{\dot P_{2\to 1}}= \frac{\dot P_{2\to 1}({\rm induced})}{\dot P_{2\to 1}({\rm induced})+ \dot P_{2\to 1}({\rm spontaneous})}=\frac{\frac{\Delta E}{2\pi} \frac{1}{e^{2\pi\Delta E/a}-1}} {\frac{\Delta E}{2\pi} \frac{1}{e^{2\pi\Delta E/a}-1} + \frac{\Delta E}{2\pi}}= e^{-2\pi\Delta E/a} \ . \ee
According to Einstein's detailed balance relation for systems in thermal equilibrium,
\be N_2/N_1= e^{-\Delta E/T}=  \frac{\dot P_{1\to 2}}{\dot P_{2\to 1}} \ , \label{eq:thermal0}\ee
the result (\ref{eq:thermal}) shows that the detector internal energy states are populated as if they were immersed in a thermal bath at the temperature $T=a/2\pi$. Therefore, following Einstein, the mean particle number per mode of the scalar radiation field should obey Planck's law
\be n_w= \frac{1}{e^{w/T}-1} \ , \ee
in agreement with the result obtained from the Bogolubov coefficient approach.

An important comment is now in order. If one considers the detector's emission and absorption rates for a final state having momentum $\vec{k}$ of the emitted scalar particle, the thermal condition (\ref{eq:thermal0}) is still satisfied for each individual mode $\vec{k}$.
This can be seen from Eqs. (\ref{rate21k}) and (\ref{rate12k}). Since the Bessel function  $K_{i2\Delta E/a }(x)$ is real and invariant under the change $\Delta E \to -\Delta E$, the ratios ${\dot P_{1\to 2,\vec{k}}}/{\dot P_{2\to 1,\vec{k}}}$ lead to the Boltzmann thermal factor $e^{-2\pi\Delta E/a}$ for every $\vec{k}$. From this observation, the result (\ref{eq:thermal}) can be easily understood, since $\dot P_{2\to 1}=e^{\pi\Delta E/a} M(\Delta E/a)$ and $\dot P_{1\to 2}=e^{-\pi\Delta E/a} M(\Delta E/a)$, where $M(\Delta E/a)$ represents the integral over momenta $\vec{k}$ of the Bessel function times the (momentum independent) common factor $g^2 |\langle E_1|m(0)| E_2\rangle|^2/(4\pi^3a)$. The thermal result (\ref{eq:thermal}), therefore, stems from the thermal properties of the transition rates to individual $\vec{k}$-modes and is not the result of integrating over all the momenta $\vec{k}$. Indeed, one may also relate the absorption and emission probability rates by considering a shift of the variable $\Delta \tau$ in Eq. (\ref{eq:time_int}) of the form $\Delta \tau\to \Delta \tau+2\pi i/a$, which immediately leads to $P_{1\to 2,\vec{k}}=e^{-2\pi\Delta E/a}P_{2\to 1,\vec{k}}$. We will come back to this point in section V.

\section{Transition probabilities and two-point functions}\label{sec:IV}

It is common in the literature (see, for instance, \cite{birrel-davies}) to express the transition probabilities computed in section \ref{sec:III} in terms of the two-point correlation function of the field. The sum over all possible  one-particle states needed to obtain the transition probabilities in the previous section, leads to a sum in modes $\sum_{\vec{k}}u^M_{\vec{k}}(x_1)u^{M*}_{\vec{k}}(x_2)$ that gives rise to the two-point function for the Minkowski vacuum. Then, if we perform the integration over all the final states in the expressions for the transition probabilities in the previous section prior to the integration in time, we get

\be P_{i \to f} = g^2 |\langle E_f |m(0)|E_i\rangle|^2 F_{i\to f}(\Delta E) \ , \ee
where
$F_{i \to f}(\Delta E)$ is the so-called response function
\be \label{FE1}F_{i \to f}(\Delta E)=
\int_{-\infty}^{+\infty}d\tau_1d\tau_2
e^{i (E_i-E_f) \Delta \tau} G_M(\Delta \tau -i\epsilon) \ ,  \ee and $\Delta \tau= \tau_1 -\tau_2$ (from now on the limit $\epsilon\to 0^+$ is understood). In the previous expressions we can have $i=1$, $f=2$ or $i=2$, $f=1$, and $\Delta E$ is positive by definition.
The quantity (\ref{FE1}) is essentially given by the Fourier transform of the Wightman two-point function $G_M(\Delta \tau -i\epsilon)$ evaluated on the detector's trajectory. For a massless field, the Wightman two-point function for the Minkowski vacuum $|0_M\rangle$ in (\ref{FE1}) is given by
\be \label{GM12}G_M(x_1, x_2)\equiv \langle 0_M|
\Phi(x_1)\Phi(x_2)|0_M\rangle= -\frac{1}{4\pi^2[(t_1-t_2 -i\epsilon)^2 -(\vec{x}_1-\vec{x}_2)^2]}  \ , \ee with $(t,\vec{x})$ inertial coordinates. The projection on the accelerated trajectory (\ref{accelerated}) gives
\be G_M(\Delta \tau -i\epsilon)= \frac{-({a}/{2})^2}{4\pi^2\sinh^2
\left[ \frac{a}{2}( \Delta \tau
-i\epsilon)\right]} \ . \label{eq:sinh}\ee
The
transition probability per unit proper time is then given by
\be \label{eq:emission} {\dot P}_{i \to f} = g^2 |\langle E_f |m(0)|E_i\rangle|^2 {\dot F_{i\to f}}(\Delta E) \ , \ee
where
\be \label{RRF}
{\dot F_{i \to f}(\Delta E)} = \int_{-\infty}^{+\infty}d\Delta
\tau e^{i (E_i -E_f) \Delta \tau}G_M(\Delta \tau -i\epsilon)  \ . \ee

Paralleling the previous section, we want to obtain separate expressions for the different processes.
For the induced emission, we can obtain an expression simply by subtracting the spontaneous emission rate (\ref{spontaneousrate}) from the total probability rate  ${\dot P}_{2 \to 1}$ given by the above expression (\ref{eq:emission})

 \be {\dot P}_{2 \to 1}({\rm induced})= g^2 |\langle E_1 |m(0)|E_2\rangle|^2 \left( {\dot F_{2\to 1}}(\Delta E)- \frac{\Delta E}{2\pi}\right) \ . \label{ind}  \ee
If we now take into account the identity
\be \label{identity}\int_{-\infty}^{+\infty}d\Delta
\tau \ e^{i\Delta E \Delta \tau} \ G_R(\Delta \tau -i\epsilon)=\frac{\Delta E}{2\pi} \ , \ee
where $G_R$ is the  vacuum two-point function of the accelerated observer (we recall that $|0_R\rangle$ is the Rindler vacuum)
\be G_R(\Delta \tau -i\epsilon)\equiv \langle 0_R|
\Phi(x_1)\Phi(x_2)|0_R\rangle= \frac{-1}{4\pi^2
( \Delta \tau -i\epsilon)^2} \ , \label{eq:GA}\ee
then expression (\ref{ind}) can be easily rewritten in terms of an integral as\footnote{Note that ${\dot P}_{2 \to 1}({\rm induced})\neq {\dot P}_{2 \to 1}$. This fact was overlooked in \cite{campo} leading to a missunderstanding of the role of the subtraction term  $G_R(\Delta \tau -i\epsilon)$ in (\ref{induced}).} 
 \be \label{induced}{\dot P}_{2 \to 1}({\rm induced})= g^2 |\langle E_1 |m(0)|E_2\rangle|^2 \int_{-\infty}^{+\infty}d\Delta
\tau e^{i\Delta E \Delta \tau}[G_M(\Delta \tau -i\epsilon)- G_R(\Delta \tau -i\epsilon)] \ . \ee

An advantage of this expression for the purely induced emission rate is that the integrand is now a smooth function over the real axis even in the absence of the $i\epsilon$. This is so because of the universal short-distance behavior of the two-point functions for any physical state [the so called Hadamard condition (see, for instance, \cite{kay-wald})] that makes the divergences of both two-point functions cancel out. Therefore, the $i\epsilon$-prescription in the integrand of (\ref{induced}) is redundant and can be omitted
\be \label{induced2}{\dot P}_{2 \to 1}({\rm induced})= g^2 |\langle E_1 |m(0)|E_2\rangle|^2 \int_{-\infty}^{+\infty}d\Delta
\tau e^{i\Delta E \Delta \tau}[G_M(\Delta \tau )- G_R(\Delta \tau )] \ . \ee
The result of this integral is
\be  \label{EFif2} \int_{-\infty}^{+\infty}d\Delta
\tau e^{i\Delta E \Delta \tau}\left[\frac{-({a}/{2})^2}{4\pi^2\sinh^2
\left[ \frac{a}{2}\Delta \tau \right]}+ \frac{1}{4\pi^2( \Delta \tau)^2} \right] = \frac{\Delta E}{2\pi} \frac{1}{e^{2\pi\Delta E/a}-1} \ , \ee
and,
therefore, the result for $\dot P_{2 \to 1} ({\rm induced})$ coincides with (\ref{inducedrate}).
\\

The absorption probability rate $\dot P_{1 \to 2}$ can be calculated in the same way, but with $\Delta E\to - \Delta E$ (recall that $\Delta E>0$, by definition). Therefore, taking into account that the integral in (\ref{EFif2}) is real, one obtains
\bea \label{excited}{\dot P}_{1 \to 2}&=& g^2 |\langle E_1 |m(0)|E_2\rangle|^2 \int_{-\infty}^{+\infty}d\Delta
\tau e^{-i\Delta E \Delta \tau}[G_M(\Delta \tau )- G_A(\Delta \tau )]\nonumber \\ &=&  g^2 |\langle E_f |m(0)|E_i\rangle|^2 \frac{\Delta E}{2\pi} \frac{1}{e^{2\pi \Delta E/a}-1}={\dot P}_{2 \to 1}({\rm induced}) \ . \eea
Thus, as expected, the absorption rate coincides with the rate of induced emission.
We reproduce in this way the same results as in the previous section.\\

It is interesting to note how the $i\epsilon$ prescription, that provided a well defined distributional sense to the two-point functions, allows one to write a single expressions (\ref{eq:emission}) to account for both the total emission and absorption probability rates. This is so because, on the one hand, the spontaneous emission probability rate can be computed as the residue of the pole of the two-point function $G_M(\Delta \tau)$ at $\Delta \tau=0$ and, on the other hand, the stimulated emission and the absorption probability rate can be computed as the sum of the residues of the infinite number of poles of $e^{i (E_i-E_f) \Delta \tau} G_M(\Delta \tau)$ in the positive and negative imaginary axes, respectively, excluding the one at $\Delta \tau=0$. The $i\epsilon$ displaces the real pole of $G_M(\Delta \tau)$ from $\Delta \tau=0$ to $\Delta \tau=i\epsilon$ (recall that $\epsilon>0$). Therefore, including the $i\epsilon$ and using the Cauchy theorem, the integral (\ref{RRF})  includes the induced and spontaneous contribution when $E_i-E_f>0$ (emission) and only the induced one when $E_i-E_f<0$ (absorption), allowing one to obtain the well known compact integral representation for both the total emission and absorption probability rates. We will come back to this point in section V, where we show that there is an advantage in writing down the separate expressions above for the different processes when one studies the influence of trans-Planckian physics.

\section{The role of trans-Planckian physics }

The thermal spectrum obtained  in the analysis in terms of Bogolubov coefficients (section \ref{sec:Rindler}) seems to depend
crucially on the exact validity of relativistic field theory on all
scales. The intermediate integral (\ref{integralkX}) involves an unbounded integration
in arbitrarily large Minkowskian momentum $k'_x$. If one introduces
an ultraviolet cutoff $\Lambda$ for $|k'_x|$ in that integral, which particularizes a given Lorentz frame,
the resulting spectrum is dramatically modified. This can be implemented, for instance, by introducing a damping factor such as $e^{-(y/2R)^2}$ [with $R\approx\ln(\Lambda/a$)] in (\ref{integralkY}) while keeping the integration limits up to infinity. Then the delta function turns into $\delta_\sigma(w_1-w_2)=e^{-(w_1-w_2)^2/\sigma^2}/\sigma\sqrt{\pi}$, where $\sigma=1/R$. The number of particles described by wave packets of the (standard) form $u_{w_jn\vec{k}_{\bot}}=\epsilon^{-1/2}\int_{w_j-\epsilon/2}^{w_j+\epsilon/2} dw e^{i 2\pi n w/\epsilon}u^{R}_{w,\vec{k}_{\bot}}$, which are localized at the instant $t_n=2\pi n/\epsilon$ and are peaked at the frequency $w_j=(j+1/2)\epsilon$ with width $\epsilon$, then becomes $\delta (\vec{k}_{\bot_1}-\vec{k}_{\bot _2})(e^{2\pi w_j/a}-1)^{-1} e^{-(t_n\sigma/2)^2}$, which decays exponentially with time as $t_n\to \infty$ showing that thermality becomes a transient process even if $\Lambda$ is at the Planck scale.\footnote{As we will show later in this section, a different conclusion is reached when we introduce the cut-off in the Bogolubov transformation method by means of a Lorentz invariant procedure.}

This apparent sensitivity of the thermal distribution of Rindler quanta to the high frequency band of the spectrum of fluctuations of the field in the Minkowski vacuum contrasts with the derivation in terms of transition probabilities of the detector of section \ref{sec:III}. To show that the detector energy levels are thermally populated (as if it were immersed in a thermal bath) does not require one to integrate over all the Minkowskian momenta. This follows from the fact that the individual transition rates for each (Minkowskian) mode of momentum $\vec{k}$ satisfy the detailed balance relation as discussed in the last paragraph of section \ref{sec:III}. In addition, it can be shown that the relative contribution of trans-Planckian Minkowski modes to the integral (\ref{crucialintegral}) is negligible, even at late times. This indicates that the thermal properties of the radiation bath that excites the detector are not crucially linked to the integration over large frequencies/momenta in  the spectrum of (real) Minkowskian modes.

Since the sum over momenta in section \ref{sec:III} is not the reason for the existence of the thermal properties, the only place where trans-Planckian physics could play a role is in the integration in $\Delta \tau$. In order to study this integral it is convenient to first integrate in $\vec{k}$, which leads to the derivation of the acceleration radiation in terms of two-point functions presented in section \ref{sec:IV}. Using (\ref{eq:emission}) and (\ref{RRF}) we have the following integral expression for the emission probability rate
\bea \label{induced3}{\dot P}_{2 \to 1}&=& g^2 |\langle E_1 |m(0)|E_2\rangle|^2 \int_{-\infty}^{+\infty}d\Delta
\tau e^{i\Delta E \Delta \tau/\hbar}[G_M(\Delta \tau-i \epsilon) ] \nonumber \\ &=& g^2 |\langle E_1 |m(0)|E_2\rangle|^2 \int_{-\infty}^{+\infty}d\Delta
\tau e^{i\Delta E \Delta \tau/\hbar}[-\frac{\hbar({a}/{2})^2}{4\pi^2\sinh^2
\left[ \frac{a}{2}(\Delta \tau-i \epsilon) \right]}]\nonumber \\  &=&g^2 |\langle E_1|m(0)| E_2\rangle|^2\frac{\Delta E}{2\pi}\frac{e^{2\pi \Delta E/a}}{e^{2\pi \Delta E/a}-1}  ,\eea
and for the absorption probability rate [see (\ref{excited})]
\bea \label{excited3}{\dot P}_{1 \to 2}&=& g^2 |\langle E_1 |m(0)|E_2\rangle|^2 \int_{-\infty}^{+\infty}d\Delta
\tau e^{-i\Delta E \Delta \tau/\hbar}[G_M(\Delta \tau-i \epsilon) ] \nonumber \\ &=& g^2 |\langle E_1 |m(0)|E_2\rangle|^2 \int_{-\infty}^{+\infty}d\Delta
\tau e^{-i\Delta E \Delta \tau/\hbar}[-\frac{\hbar({a}/{2})^2}{4\pi^2\sinh^2
\left[ \frac{a}{2}(\Delta \tau-i \epsilon) \right]}]\nonumber \\  &=&g^2 |\langle E_1|m(0)| E_2\rangle|^2\frac{\Delta E}{2\pi}\frac{1}{e^{2\pi \Delta E/a}-1}  . \eea
In the above integrals, the $i\epsilon$ term plays a fundamental role in regularizing the denominator of the integrand, which otherwise would lead to a divergence as $\Delta\tau\to 0$. This could be seen as an indication that ultra-short (sub-Planckian) distances $(\Delta \tau)^2 < \ell_P^2$ ($\ell_P$ is the Planck length) play a relevant role in the outcome of those integrals. A Planckian cutoff for $\Delta \tau$ in the above integrals substantially  modifies the thermal result. However,  the integral representation for the transition probability rates provided by the $i\epsilon$ prescription cannot be properly used to evaluate the effect of such a cut-off. The $i\epsilon$ prescription is incompatible with cutting out part of the integration path \cite{kay-wald}.  The distributional character of the integrand, in contrast to the smooth integrand of (\ref{EFif2}),  prevents us from properly evaluating  the relative contribution of  trans-Planckian physics in terms of the above integral expressions.

On the contrary, when the different contributions  to the transition processes are worked out separately  (see expressions (\ref{induced2}) and (\ref {EFif2}) in section \ref{sec:IV}) the integrals are well defined smooth functions. This implies that expression (\ref{induced2}), as pointed out in \cite{agullo-navarro-olmo-parker08a, agullo-navarro-olmo-parker08b}, can be used to properly estimate the interval of $\Delta \tau$ that significantly contributes to the overall integral for the induced emission rate
\be \label{int}  \int_{-\infty}^{+\infty}d\Delta
\tau e^{-i\Delta E \Delta \tau}\left[\frac{-({a}/{2})^2}{4\pi^2\sinh^2
\left[ \frac{a}{2}\Delta \tau \right]}+ \frac{1}{4\pi^2( \Delta \tau)^2} \right] = \frac{\Delta E}{2\pi} \frac{1}{e^{2\pi\Delta E/a}-1} \ . \ee
One finds that values of $\Delta \tau$ giving the dominant contribution to the above integral are of the same order as the acceleration $a$ itself (for non-extreme values of $\Delta E$). That is, neither very large nor very small values of $\Delta \tau$, in comparison with $a$, are important for obtaining the result. To be more precise, one can compute the integral (\ref{int}), excluding the contribution of ultra-short proper time lapses $|\Delta \tau| < \ell_P$, and the result for the induced emission probability rate is
\be \dot P_{2 \to 1} ({\rm induced})=  g^2 |\langle E_1 |m(0)|E_1\rangle|^2 \frac{\Delta E}{2\pi}\left( \frac{1}{e^{2\pi\Delta E/a}-1}-\frac{a\, \ell_P}{48 \pi^3 \Delta E/a}+O(a \ell_P)^3\right) \ , \ee
where we can see that the correction term is completely negligible relative to the thermal term if $a \ll \ell_P^{-1}$ and if the energy gap $\Delta E$ is not much larger than the temperature, $a/(2\pi)$, of the thermal spectrum. Exactly the same result is obtained  for the excitation probability rate. Obviously, the spontaneous emission is robust against trans-Planckian physics.\footnote{The result
 $\dot P_{2 \to 1}({\rm spontaneous})=  g^2 |\langle E_1|m(0)| E_2\rangle|^2  \frac{E_2 - E_1}{2\pi}$
stems essentially from evaluating the amplitude of the modes defining the vacuum at the (low) frequency $w=\Delta E\equiv E_2 - E_1$. Moreover, the derivation in terms of the accelerated detector indicates that the integral in (\ref{A21integral})
does not require very large $|\vec{k}_{\bot}|$ due to the exponential decay of the Bessel function.
Additionally, the spontaneous emission rates of all microscopic systems calculated in the conventional low-energy frameworks agree with observations.}
Therefore, we conclude that for an accelerated detector, the behavior of the two-point function relative to Planckian lapses of proper time does not affect the bulk of the thermal radiation.

The above  analysis indicates that the spectrum of thermal radiation felt by a uniformly accelerated observer in Minkowski spacetime is rooted on energy scales of the same order as the energy scale defined by the temperature of the spectrum. We want to point out that the derivation of the transition probability rates in terms of two point functions in the previous section, has allowed us to introduce a cut-off in such a way that Lorentz invariance is manifestly respected. This contrasts with the introduction of a non-Lorentz-invariant cut-off as is sometimes felt to be necessary to avoid appealing to trans-Planckian frequencies in computing the Bogolubov coefficients.  Such a cut-off which distinguishes a particular inertial frame also produces a substantial modification of the thermal spectrum at late times. This late time modification is an explicit consequence of the break down of Lorentz invariance, because different instants along the trajectory are related by a Lorentz boost. However, when defined in a Lorentz invariant way, as in our analysis, trans-Planckian physics does not play a fundamental role in obtaining the thermal spectrum of the acceleration radiation.\\

In the remaining part of this section, we want to extend the previous analysis of the trans-Planckian contribution to the thermal spectrum, as measured by a particle detector, to the method of calculating the mean number per mode of  ``Rindler'' particles $n_w$ present in the Minkowski vacuum state. The derivation of this mean number using Bogolubov coefficients has already been summarized in section \ref{sec:Rindler}. We will also make some comments about the extension of our analysis to the computation of Hawking radiation by black holes.

In the first paragraph of this section, we showed that introducing a high frequency cut-off in the calculation of Bogolubov coefficients strongly affects the thermal spectrum. This contrasts with our previous conclusion.
This apparent conflict boils down to whether or not one insists on respecting Lorentz invariance. We prefer to preserve this symmetry.
To achieve this, we reformulate the analysis of the mean number distribution of quanta obtained in section \ref{sec:Rindler} in such a way that a study of the trans-Planckian contribution to the thermal spectrum can be done using invariant quantities, thus paralleling our analysis in terms of particle detectors.

As originally  shown by Fulling \cite{fulling}, the spectrum of acceleration radiation can be derived by computing the content of ``Rindler'' particles in the Minkowski vacuum state.  This mean number of Rindler particles per mode $n_w$ can be expressed as $n_w\equiv \langle 0_M| N_w^{R}| 0_M \rangle$, where $N_w^{R}$ is the Rindler particle number operator.  In terms of Bogolubov coefficients, that quantity is evaluated as
\be \label{Nbogolubov} n_w=\langle 0_{M} |N^{R}_w| 0_{M}\rangle =\sum_{w'}|\beta_{ww'}|^2 \ . \ee
On the other hand, as explained in \cite{agullo-navarro-olmo-parker07, agullo-navarro-olmo-parker08a, agullo-navarro-olmo-parker08b}, one can rewrite the previous expression in terms of two-point functions as
\be\label{eq:N-epsFH}
\langle 0_{M} |N^{R}_w| 0_{M}\rangle = \int_\Sigma
d\Sigma_1 ^\mu d\Sigma_2 ^\nu
[u^{R}_{w,\vec{k}_{\bot}}(x_1){\buildrel\leftrightarrow\over{\partial}}_\mu
][u^{R*}_{w,\vec{k}_{\bot}}(x_2){\buildrel\leftrightarrow\over{\partial}}_\nu
] (G_{M}(x_1, x_2)- G_{R}(x_1, x_2))\ , \ee
where $\Sigma$ is a Cauchy hypersurface, $u^{R}_{w,\vec{k}_{\bot}}(x)$ are the Rindler modes defined in (\ref{modesR}) and $G_{M}, G_{R}$ are the two-point functions for the Minkowski and Rindler vacuum states, respectively.
Choosing the null plane $H^-$, defined by $V\equiv t+x =0$, as the initial data hypersurface, we obtain\footnote{We neglect an (infinite) factor $ \delta( 0 )$ arising in the integral as a consequence of using plane waves modes. By using the standard normalizable wave-packets that factor disappears.}
\be \label{nwr}n_w=\langle 0_M| N_w^{R}| 0_M \rangle =  \frac{2\pi}{w}\int_{-\infty}^{+\infty}d\Delta
u \ e^{-iw \Delta u}\left[\frac{-({a}/{2})^2}{4\pi^2\sinh^2
\left[ \frac{a}{2}\Delta u\right]}+ \frac{1}{4\pi^2( \Delta u)^2} \right] = \frac{1}{e^{2\pi w/a}-1}  \ , \ee
where $u$ is the null coordinate $u\equiv \tau -\xi$ and $\Delta u\equiv u_1-u_2$. Note in passing that, if we project the acceleration trajectory ($\xi=0$) onto the horizon $H^-$, then the  point on $H^-$ characterized by the coordinate $u$  corresponds to the point on the uniformly accelerated trajectory characterized by coordinate $\tau$.

We want to point out now that the previous derivation of the thermal spectrum using equations (\ref{eq:N-epsFH}) and (\ref{nwr}) is closely related with the derivation presented in section \ref{sec:IV} using particle detectors and two-point functions. To be more precise, if we compare the generic relation $\dot P_{1\to 2}= Bu_w$ (recalling that $u_w$ represents the energy density per mode $w$ of the radiation) with the  result for the excitation emission rate (\ref{excited}), we see
that the mean number of particles per mode $w$ for the thermal distribution corresponds to the integral
\be \label{nwd}n_w\equiv \frac{1}{e^{2\pi w/a}-1}= \frac{2\pi}{w}\int_{-\infty}^{+\infty}d\Delta
\tau e^{-iw \Delta \tau}\left[\frac{-({a}/{2})^2}{4\pi^2\sinh^2
\left[ \frac{a}{2}\Delta \tau \right]}+ \frac{1}{4\pi^2( \Delta \tau)^2}  \right]   \ , \ee
This integral coincides with (\ref{nwr}), with $\Delta \tau$ replaced by $\Delta u$. Note that along $H^-$, the quantity $\Delta u$ is invariant under Lorentz transformation.
Thus, we see that there is a clear relation between the derivation of the acceleration radiation using accelerated particle detectors and the derivation based on the Rindler particle number.
The former derivation showed that invariantly defined trans-Planckian physics does not significantly affect the observed radiation. This implies that we use the condition $|\Delta u|< \ell_P$ to characterize in a Lorentz invariant way the trans-Planckian physics in equations (\ref{eq:N-epsFH}) and (\ref{nwr}). A further discussion of the Lorentz-invariant cut-off introduced here and the comparison with the cut-off $|\Delta U|< \ell_P$, with $U\equiv t-x$,  can be found in \cite{agullo-navarro-olmo-parker09, helfer}.  \\

The above discussion offers some hints for the study of the trans-Planckian question in Hawking radiation by black holes.
For a spherically symmetric black hole, the average number of particles observed at late times
in the state in which no particles are present at early times is given by an expression analogous to (\ref{Nbogolubov}), but in the black hole geometry \cite{hawking}. A steady rate of radiation is obtained from an explicit computation of the corresponding Bogolubov coefficients  and it
turns out to be thermal. However, to get this one
needs to perform an unbounded integration in the frequencies $w'$, as discussed, for example, in
\cite{Parker77, Wald84, jacobson9193, fabbri},
in parallel to the unbounded integration in $k'_X$ for the acceleration radiation effect in (\ref{integralkX}). A cutoff in the frequencies $w'$ will change completely the
Hawking effect.
It will introduce a damping time-dependent factor modulating the thermal radiation. The Hawking radiation is then converted into
a transient phenomenon (see, for instance, \cite{agullo-navarro-olmo-parker07} and also \cite{gil}).

In analogy with the acceleration radiation effect, it is possible to derive the Hawking effect in terms of smooth integrals involving the difference of the two-point functions of the two vacuum states involved. In fact, the general expression (\ref{eq:N-epsFH}) can also be applied to the black hole, with the Minkowski observer replaced by the so-called {\it in} observer, the Rindler observer by the {\it out} observer, and the acceleration $a$ by the surface gravity $\kappa$ of the black hole (for details see  \cite{agullo-navarro-olmo-parker07, agullo-navarro-olmo-parker09,agullo-navarro-olmo-parkerkerr}).  Exactly the same expression (\ref{nwr}) is then obtained,\footnote{The analogous of expression (\ref{nwr}) for black holes is also relevant \cite{agullo-navarro-olmo-parker07, agullo-navarro-olmo-parker08b,agullo-navarro-olmo-parkerkerr} to preserve the near-horizon two-dimensional conformal symmetry of black holes, which seems to play a crucial role for the understanding of the Bekenstein-Hawking entropy (see, for instance, \cite{carlip, skenderis, navarro, agullo}) .} where now $u$ stands for the retarded null coordinate $u\equiv t - r^*$, with $t$ the Schwarzschild time and $r^*$ the tortoise coordinate.
The analysis performed for the acceleration radiation, then suggests that (as was done in \cite{agullo-navarro-olmo-parker07, agullo-navarro-olmo-parker08b}) the condition $|\Delta u|<\ell_P$ characterizes the regime of trans-Planckian physics entering into the derivation of the thermal spectrum and that altering physics in this trans-Planckian regime will not modify the fundamental properties of the Hawking radiation.

\section{Conclusions}

In this work we have analyzed the trans-Planckian question for a uniformly accelerated detector. We have split the transition probability rates into spontaneous and induced contributions. The latter can be expressed as a Fourier integral with a smooth integrand involving the difference of two-point functions. This permits us to estimate in a new way the contribution of trans-Planckian physics to the induced probability rates and allows us to  show that the main contribution to the induced rates  comes from the low energy scale defined by the acceleration $a$. Trans-Planckian (and ultralow energy) contributions seem not to play a central role. Nevertheless, one cannot discard that new effects could arise at the Planck scale if one admits that at such high energies non-linear couplings of the field and detector emerge or, even more, if the very notion of the spacetime and Lorentz invariance dissolve into more elementary structures. In other words, we have assumed the validity of the field-detector model up to energies well above the natural scale of the system.
On the other hand,  the close analogy between the acceleration radiation and Hawking effect suggests that the above arguments also support the view of  the Hawking effect as a low energy phenomenon, in agreement with  recent results coming from a different perspective \cite{Balbinot}.

\section*{Appendix A: Spontaneous emission of a  detector in de Sitter space}

 In this appendix we evaluate the spontaneous emission rate of the detector in a de Sitter space described by the static metric $ds^2= -(1-{\tilde r}^2H^2)d\tilde t^2+ (1-{\tilde r}^2H^2)^{-1}d\tilde r^2 +\tilde r^2d\Omega^2$. To properly compare this emission rate  with that of the massless (conformal) field in Minkowski space analyzed in section II we have to consider here a massless field with a conformal coupling $\xi=1/6$ to the curvature. In this situation the form of the modes $u_{wlm}(\tilde t, \tilde r, \theta, \phi)$ on the detector's trajectory $\tilde t =\tau, \tilde r=0$ (detector at rest and at the origin of static coordinates) is\footnote{At an arbitrary point the $s$-wave mode is $u_{wl=0}(\tilde t, \tilde r) = \tilde r^{-1}(\pi w)^{-1/2}e^{-iw\tilde t}\sin w(H^{-1}\tanh^{-1}(H\tilde r))Y_{00}$.}
\be u_{wlm}(\tau) = \sqrt{\frac{w}{\pi}}e^{-iw\tau}\frac{1}{\sqrt{4\pi}}\delta_{l0} \ . \ee
The transition probability for all possible one-particle final states is given by
\bea \label{A21integraldS}\dot P_{2 \to 1}({\rm spontaneous}) &=&g^2 |\langle E_f|m(0)| E_i\rangle|^2 \sum_{lm}\int_0^{+\infty} dw \frac{w}{2\pi}\delta (E_1 -E_2 +w)\delta_{l0} \nonumber \\ &=& g^2 |\langle E_1|m(0)| E_2\rangle|^2  \frac{(E_2 - E_1)}{2\pi}\ , \eea
which, as expected, coincides with (\ref{A21f}).\\

\noindent { \bf Acknowledgements.} This
work has been partially supported by the Spanish grants FIS2008-06078-C03-02 and  the  Consolider-Ingenio 2010 Programme CPAN (CSD2007-00042). I.A. and L.P.  have been partly
supported by NSF grants PHY-0071044 and PHY-0503366 and by
a UWM RGI grant.  G.O.
thanks MICINN for a JdC contract and the ``Jos\'e Castillejo'' program for funding a stay at the University of Wisconsin-Milwaukee.\\


\begin{thebibliography}{99}


\bibitem{parker} L. Parker, {\it Phys.Rev.Lett.} {\bf 21} 562 (1968); {\it Phys. Rev.} {\bf 183}, 1057(1969).

\bibitem{fulling} S.A. Fulling, {\it Phys. Rev. D} {\bf 7}, 2850 (1973).

\bibitem{davies} P.C.W. Davies, {\it J. Phys. A} {\bf 8}, 609 (1975).

\bibitem{hawking}
S.W. Hawking, {\it Comm. Math. Phys.} {\bf 43}, 199 (1975).

 \bibitem{parker-toms}  L. Parker L. and D.J. Toms, {\it Quantum field theory in curved spacetime: quantized fields
and gravity}, Cambridge University Press, (2009).

\bibitem{unruh} W.G. Unruh, {\it Phys. Rev.} D {\bf 14} 870 (1976).

\bibitem{crispino}  L.C.B. Crispino, A. Higuchi and G.E.A. Matsas,
 {\it Rev. Mod. Phys.} {\bf 80}, 787 (2008).

\bibitem{fredenhagen-haag} K. Fredenhagen and R. Haag, {\it
Commun. Math. Phys.} {\bf 127} 273 (1990)

\bibitem{Parker77}
L.Parker, in ``Asymptotic structure of space-time'', ed. by F.P.Esposito and L.Witten, Plenum Press,N.Y.(1977)
\bibitem{Wald84} R.M. Wald {\it General Relativity}, Chicago University Press (1984)
\bibitem{jacobson9193}
T. Jacobson, {\it Phys. Rev.} D {\bf 44}
1731 (1991); {\it Phys. Rev} D {\bf 48} 728 (1993)
\bibitem{dispersionrelations}W.G. Unruh, {\it Phys. Rev.} D {\bf 51} 2827 (1995).
R. Brout, S. Massar, R. Parentani and P. Spindel,{\it Phys. Rev.} D {\bf 52} 4559 (1995).
S. Corley and T. Jacobson, {\it Phys. Rev.} D {\bf 54} 1568 (1996).
{\it Phys.Rev.} D {\bf 59} 124011 (1999).
S. Corley, {\it Phys. Rev.} D {\bf 57} 6280 (1998).
R. Balbinot, A. Fabbri, S. Fagnocchi and R. Parentani, Riv. Nuovo Cimento {\bf 28}, 1 (2005). R. Schützhold and W.G. Unruh, {\it Phys. Rev. D } {\bf 78}, 041504 (2008). M. Rinaldi {\it Phys. Rev. D}  {\bf 77}, 124029 (2008).

\bibitem{wald01} R.M. Wald, {\it Living Rev. Rel.} {\bf 4}, 6 (2001).

\bibitem{unruh-wald} W.G. Unruh and R.M. Wald, {\it Phys. Rev. D} {\bf 29}, 1047 (1984).

\bibitem{dewitt} B.S. DeWitt, in {\it General Relativity}, eds. S.W. Hawking and W. Israel, Cambridge University Press, (1979).

\bibitem{birrel-davies} N.D. Birrel and P.C.W. Davies, {\it Quantum fields in curved space}, Cambridge University Press, (1982).
\bibitem{campo} D. Campo, arXiv:1004.5324.


\bibitem{kay-wald} B.S. Kay and R.M. Wald, {\it Phys. Rep. } {\bf 207}, 49 (1991).

\bibitem{agullo-navarro-olmo-parker08a} I. Agullo, J. Navarro-Salas, G.J. Olmo and
L. Parker, {\it Phys. Rev.} D {\bf 77}, 104034 (2008)
\bibitem{agullo-navarro-olmo-parker08b} I. Agullo, J. Navarro-Salas, G.J. Olmo and
L. Parker,{\it Phys. Rev.} D {\bf 77}, 124032 (2008).

\bibitem{agullo-navarro-olmo-parker07} I. Agullo, J. Navarro-Salas, G.J. Olmo and
L. Parker, {\it Phys. Rev. D}  {\bf 76} 044018 (2007).  I. Agullo, J. Navarro-Salas, G.J. Olmo, {\it Phys. Rev. Lett.} {\bf 97}, 041302 (2006).

\bibitem{agullo-navarro-olmo-parker09} I. Agullo, J. Navarro-Salas, G.J. Olmo and
L. Parker, {\it Phys. Rev. D}  {\bf 80}, 047503 (2009); {\it Phys. Rev. D}  {\bf 81}, 108502 (2010).
\bibitem{helfer} A. Helfer {\it Phys. Rev.} D {\bf 81}, 108501(2010).

\bibitem{fabbri} A. Fabbri and J. Navarro-Salas, {\it Modeling black hole evaporation}, ICP-World Scientific, (2005).



\bibitem{gil}C. Barcelo, L.J. Garay and G. Jannes, {\it Phys. Rev.} D {\bf 79}, 024016 (2009).

\bibitem{agullo-navarro-olmo-parkerkerr} I. Agullo, J. Navarro-Salas, G.J. Olmo and
L. Parker, arXiv:1006.4404 (to appear in Phys. Rev. Lett.).

\bibitem{carlip} S. Carlip, {\it Gen. Rel. Grav.} {\bf 39}, 1519 (2007); {\it Phys. Rev. Lett.} {\bf 82}.
2828 (1999).
\bibitem{skenderis} K. Skenderis, Lect. Notes Phys. {\bf 541}, 325 (2000)
\bibitem{navarro} J. Navarro-Salas and P. Navarro, {\it Nucl. Phys.} B{\bf 579}, 250
(2000); D.J. Navarro, J. Navarro-Salas and P. Navarro, {\it
Nucl. Phys.} B{\bf 580}, 311 (2000)
\bibitem{agullo} I. Agullo, E. F. Borja, and J. Diaz-Polo, {\it JCAP} 0907:016 (2009).
\bibitem{Balbinot} R. Balbinot, A. Fabbri,  S. Fagnocchi, A. Recati, I. Carusotto, {\it Phys. Rev.A} {\bf 78}, 021603 (2008). I. Carusotto, S. Fagnocchi, A. Recati,  R. Balbinot,  A. Fabbri, {\it New J. Phys.} {\bf 10}, 103001 (2008).

\end{thebibliography}
\end{document}